\documentclass[seceq]{ptptex}

\usepackage[dvipdfm]{graphicx}
\title{%        %You can use \\ for explicit line-break.
The exchange fluctuation theorem in quantum mechanics%
}

\author{%       %Use \scshape for the family name.
Shiho \textsc{Akagawa}$^{1}$%$^{1,}$ 
and Naomichi \textsc{Hatano}$^{2,}$\footnote{E-mail: hatano@iis.u-tokyo.ac.jp}
}

\inst{%     %Affiliation, neglected when [addenda] or [errata].
$^1$Department of Physics, University of Tokyo, Tokyo 153-8505, Japan
\\
$^2$Institute of Industrial Science, University of Tokyo,\\ 
Tokyo 153-8505, Japan
}

\abst{%         %This abstract is neglected when [addenda] or [errata].
We study the heat transfer between two finite quantum systems initially at different temperatures. 
We find that a recently proposed fluctuation theorem for heat exchange, 
namely the exchange fluctuation theorem 
[C. Jarzynski and D. K. W\'{o}jcik, Phys. Rev. Lett. \textbf{92}, 
230602 (2004)], 
does not generally hold in the presence of a finite heat transfer as in the original form 
proved for weak coupling. 
As the coupling is weakened, 
the deviation from the theorem and the heat transfer vanish 
in the same order of the coupling. 
We then discover a condition for the exchange fluctuation theorem to hold 
in the presence of a finite heat transfer, namely the commutable-coupling condition. 

We explicitly calculate the deviation from the exchange fluctuation theorem 
as well as the heat transfer 
for  simple models. 
We confirm for the models that the deviation indeed 
has a finite value as far as the coupling between the two systems is finite except for the special point of the commutable-coupling condition. 
We also confirm analytically 
that the commutable-coupling condition indeed lets the exchange fluctuation theorem 
hold exactly under a finite heat transfer.  
}

\begin{document}

\maketitle

\section{Introduction}
%Œã'Å—ÊŽqŒn'Æ'ÌŒq'ª'è'É'à'Á'Ä'¢'Á'½•û'ª'¢'¢'Ì'Å'Í'ÆŽw"EBby Hatano@—vŒŸ"¢
%¬'³'¢Œn'Ìà–¾		
The development of the modern techniques of microscopic manipulation 
enables us to treat small systems, for example, nano-devices and molecular motors. 
%, nanoscale systems for example. 
In such small systems, classical thermodynamics
is not well applicable to quantification of the heat flow or the work. 
At the nano- and micro-scales, the available thermal energy per degree of freedom 
is comparable to the energy of the small systems. 
This thermal energy enhances fluctuations, whose effects measurably appear in such small systems. 
Thus, we cannot apply classical thermodynamics to these small systems. 
We need a substitute framework if we try to design or control nano-devices and molecular motors as macroscopic heat engines. 
%For example, molecular motors gain the energy from chemical reactions and 
%perform their functions continuously dissipating the energy into their environment. 
%Besides, they use thermal fluctuations in their operation. 
%Thus, we cannot describe the molecular motors using classical thermodynamics. 

%—h'炬'̒藝@Evans'ç'Ì2004"N'ÌŽÀŒ±'ðŽQl'É'µ'Ä
%In the last two decades, many studies have been made on the fluctuations in classical  system. 
%since the pioneering work by Evans $et \ al.$\cite{FToriginal1993:Evans}.
%

Since the pioneering work by Evans \textit{et al.},\cite{FToriginal1993:Evans} 
researchers have developed various results collectively 
known as the fluctuation theorem. 
The fluctuation theorem is roughly summarized as an equation 
which
relates the probability $p(\Omega)$ of observing the entropy increase $\Omega$, 
to the probability $p(-\Omega)$ of observing the entropy decrease of 
the same magnitude: 
%		TFT'Ì•\Œ»
\begin{align}
\frac{p(\Omega)}{p(-\Omega)}
	=
		{\rm e}^{\Omega}.
\end{align} 
The definition of the entropy increase $\Omega$ depends on the dynamics 
of the system under consideration. 
However, the fluctuation theorem has been established for several class of systems 
with reasonable definitions of $\Omega$.  
When the fluctuation theorem is applied to the transient response of a system, the theorem is referred to as the transient fluctuation theorem.\cite{FToriginal1994:Evans} 

%'»'Ì'¼'Ì—h'炬'̒藝@—ÊŽqŒn'àŠÜ'Þ
For more than 15~years, %Since the above works, 
%Since the pioneering work by Evans $et \ al.$\cite{FToriginal1993:Evans}, 
many versions of the fluctuation theorem have been presented for a variety of classical-mechanical  situations: %contexts'Å'à'¢'¢'©'à
thermostated systems,\cite{FToriginal1993:Evans,FToriginal1994:Evans,FT1995:GallavCohen} 
stochastic systems,\cite{FTStocha1998:Kurchan,FTStocha1999:Lebow} 
and externally driven systems.\cite{JE1997:Jarzyn, FTw1998:Crooks,FTw1999:Crooks} 
%ƒŒƒrƒ…['̏Љî
These classical-mechanical versions of the fluctuation theorem have been reviewed 
in Refs.\citen{RevFT2002:Evans,RevFT2008:Searles}. 
%ŽÀŒ±'ÌŒ‹‰Ê'ð"ü'ê'é'Ȃ炱'±
Some of these theorems were verified experimentally.\cite{ExpSSFT2004:Carberry,ExpJE2002:Liphardt,ExpFTw2005:Collin} 
%'Ç'ÌŽÀŒ±'ª‰½'ðŒŸØ'µ'½'©'ð''­'±'ƁBTFT'ª'Ç'±'©'ɍs'Á'Ä'¢'éB—vŒŸ"¢

%Œn'ð¬'³'­'µ'Ä'¢'­'Æ—ÊŽqŒn'É
When the system becomes further smaller, quantum effects may become significant. 
It is, however, not straightforward to extend the afore-mentioned classical-mechanical fluctuation theorem to quantum-mechanical systems. 
%'ª'è'ÌŒø‰Ê'ª—ÊŽqŒn'Å'Í' 'éB
The crucial difference of the quantum systems from the classical systems  is an essential role of measurements. 
In order to generalize the fluctuation theorems to quantum systems, we need to 
identify the entropy, the work, and the heat that are measured in the quantum-mechanical 
context.

%‰‰ŽZŽq'Å'è‹`'·'é•û–@'ÆTwo-time'ª'è
There have been two attempts to do this: 
first, defining operators to represent the heat and the work; 
second, measuring the system and using the measurement outcomes to represent the heat and the work. 
%‰‰ŽZŽq'Å'è‹`'·'é•û–@B
In general, 
the former attempt has led to quantum corrections to the classical results.\cite{OpeWork2000:Yukawa,OpeWork2003:Monnai,OpeWorkHeat:Allahverdyan,OpeWorkHeat:GelinKosov} 
%Two-timeԻՏ
In the latter attempt, on the other hand, several versions of the fluctuation theorem have been shown without quantum corrections.\cite{TwoTime:Kurchan,TwoTime:HalTasaki,TwoTime:Mukamel,TwoTime:Monnai,TwoTime:TalknerWorkNot} 
Both the heat and the work are defined as the difference between the results of two measurements, a two-point quantity. 
We refer to this attempt as the two-point measurement. 

The exchange fluctuation theorem (XFT)\cite{JarzynWo} was proposed 
in the framework of the two-point measurement 
by C. Jarzynski and D. K. W\'{o}jcik\cite{JarzynWo}. 
The theorem concerns the statistics of heat exchange 
between two finite systems of Hamiltonian dynamics, 
initially prepared  at different temperatures.
Let $\beta_A$ and  $\beta_B$ denote the inverse temperatures at which 
the systems $A$ and $B$ are prepared, respectively.
The symmetry relation of the exchange fluctuation theorem 
is expressed with the probability distribution $p_{\tau}(Q)$ of the 
net heat transfer $Q$ as follows:
%XFT'ð''«Ž¦'µ'Ä'¨'­B
\begin{align}
p_{\tau}(Q) = {\rm e}^{\Delta \beta Q} p_{\tau}(-Q),
%p_{\tau}(Q) = {\rm e}^{\Delta \beta Q} p^{\rm R}_{\tau}(-Q),
\label{XFT}
\end{align}
%	Y
where 
$\Delta \beta = \beta_B - \beta_A$ 
is the difference between the inverse temperatures 
and $\tau$ is the time duration between the two measurements. 
They referred to this symmetry relation as the exchange fluctuation theorem 
because the exchanged energy $Q$ between the two systems is regarded as a heat 
transfer. 
Equation~\eqref{XFT} implies that the average of 
${\rm e}^{-\Delta \beta Q}$ over the ensemble of realizations for any time interval $\tau$ 
is unity:
%XFT'©'ç'à'½'炳'ê'é‹AŒ‹B
\begin{align}
\langle {\rm e}^{-\Delta \beta Q} \rangle_{\tau}
%		&=\int^{\infty}_{-\infty} dQ p_{\tau}(Q) {\rm e}^{-\Delta \beta Q}
%\label{DefIXFT}		 \\
%		&= \int^{\infty}_{-\infty} dQ p^{\rm R}_{\tau}(-Q) 
%		\nonumber \\
		&=1.
%		&=\sum_{n, m} P( m | n; \tau)  {\rm e}^{-\Delta \beta Q_{m \to n}}
\label{IXFTOriginal}
\end{align}
%	Y
This integral form of equality is a direct consequence 
of the exchange fluctuation theorem, 
and thus we refer to this equality as the integral exchange fluctuation theorem (IXFT).

Jarzynski and W\'{o}jcik derived the exchange fluctuation theorem 
in both classical and quantum systems. 
We will discuss the exchange fluctuation theorem only for quantum systems 
in the present paper. 
%The fluctuation theorems represent a degree of irreversibility in finite systems. 
%Owing to the similarity to the concept of second law of thermodynamics, these theorems %are not only of basic theoretical relevance but expected to play an important role in %design of nanotechnological devices and in the understanding of biological processes.
%Ä'ÑXFT'É–ß'Á'Ä
The situation in which the exchange fluctuation theorem was ``proved''\cite{JarzynWo} 
is quite simple; 
the two finite quantum systems are prepared in equilibrium at different temperatures, 
and then placed in thermal contact with one another. 
There is no  work resource such as an external field nor an external force. %on the system, 
%and thereby 
%the energy of the each system can change as the heat only. 
In this situation, %of the exchange fluctuation theorem, 
we simply identify the energy increase of each system as a heat flowing into the system. 
%We consider that it might be helpful 
%for obtaining deep understanding of the fluctuations 
%to study quantum-mechanical systems in which we can analyze the  heat and the work 
%separately. 
Equations~\eqref{XFT} and~\eqref{IXFTOriginal} were then ``proved'' in the case 
where the coupling between the two quantum systems is weak.

%separately form the work is important to 
%understand its characteristics and differences from the work. 
%
%and it gives us deep understanding of the fluctuations of small-systems. 
% 
%Thus it is important and meaningful to study the heat transfer between the finite quantum 
%systems in order to reveal and understand characteristics of the heat transfer 
%between the small systems.
%Œ¤‹†'Å•ª'©'Á'½'±'Æ'Ì'Ü'Æ'ß
To summarize the present paper, we find the following:
\begin{enumerate}
\item %ˆê"Ê'É'ÍXFT'͐¬'è—§'½'È'¢
	The exchange fluctuation theorem does not generally hold 
	in the presence of a finite heat transfer 
	as in the original form proposed for weak coupling. 
\item %CCC‰º'Å'ÍXFT'ÍŒµ–§'ÈŽ®'Æ'È'éB
	If the Hamiltonian that couples the two systems 
	commutes with the total Hamiltonian, 
	the exchange fluctuation theorem becomes an exact relation 
	for arbitrary strength of coupling. 
	We refer to this condition as the commutable-coupling condition. 
\if0
\item %CCC‰º'Å'ÍXFT'ÍH. Tasaki'ª"±o'µ'½"MŒðŠ·'ÉŠÖ'·'éŠÖŒWŽ®'ƈê'v'·'éB
	 Under the same situation as ii., the exchange fluctuation theorem coincides with 
	 Tasaki's fluctuation theorem about heat transfer. 
	 From the view point, the exchange fluctuation theorem is just a approximation 
	 of Tasaki's fluctuation theorem. 
\fi
\end{enumerate}
%'±'̘_•¶'̍\¬%
In short, the exchange fluctuation theorem holds not because of weak coupling 
but because of the commutable-coupling condition.

The present paper is organized as follows.
%	—ÊŽqŒn'Å•ª'©'Á'Ä'¢'邱'Æ'ðŠ®'S'ɃtƒHƒ['·'é''à'è'Å'â'Á'½•û'ª'¢'¢'Ì'©?
%	—vŒŸ"¢
In \S\ref{Chap:XFT}, 
we first introduce the joint probability in the two-time measurement 
and derive a symmetry relation about the joint probability. 
Then, we show deviation from the exchange fluctuation theorem. 
%After that, 
In \S\ref{sec:CCC}, 
we introduce a condition on the coupling Hamiltonian for the exchange fluctuation theorem to hold exactly, 
which we refer to as the commutable-coupling condition. 
In \S\ref{sec:Example}, 
we show an explicit form of the deviation for simple models 
and demonstrate that the deviation from the exchange fluctuation theorem vanishes 
under the commutable-coupling condition.
Conclusions will be presented in \S\ref{Chap:Conclusion}.
%˜_•¶'̍\¬@'±'̏ã%

%%%%
\section{Exchange fluctuation theorem in quantum mechanics}
\label{Chap:XFT}
%	Joint Prob. 'Ì'è‹`B
\subsection{Joint probability}
In this section, we define a procedure of the two-point measurement 
and the corresponding joint probability. 
Assuming the time-reversal invariance, 
we derive a symmetry relation of the joint probability.  

%Jarzynski and W\'{o}jcik \cite{JarzynWo} 
We consider two finite quantum systems given by the Hamiltonian
%ˆµ'¤ƒnƒ~ƒ‹ƒgƒjƒAƒ"'Ì'è‹`
\begin{align}
H=&H_{0} + \gamma H_{\rm c},
\label{Hamiltonian}
\\
H_{0} =& H_A \otimes 1_B + 1_A \otimes H_B, 
\end{align}
where $H_{A}$ and $H_{B}$ are the Hamiltonians of the systems $A$ and $B$, 
respectively. 
The second term of the right-hand side of Eq.~\eqref{Hamiltonian}, 
$H_{\rm c}$, is the coupling Hamiltonian which describes 
the connection between the two systems and $\gamma$ is a factor 
controlling the coupling strength 
between the two systems. 
In the present paper, we consider the case where $H_{\rm c}$ does not have 
any extra degrees of freedom other than in the systems $A$ and $B$. 

Let $ |m_\alpha\rangle$ and $ E^{\alpha}_{m_{\alpha}} $ 
denote an eigenstate of $H_\alpha$ and the corresponding eigenvalue ($\alpha=A, B$), 
respectively. 
We refer to the product states $|m_A\rangle \otimes |m_B\rangle$ as $|m\rangle$ 
for simplicity. 
The Hilbert space is spanned by the set of $|m\rangle$. 

%'Q"_'ª'è'Ìà–¾
Here, we introduce the measurement procedure to discuss the heat exchange between 
the two systems over a finite time duration $0\leq t \leq \tau$. 
%'ª'è'͎ˉe'ª'è
We consider a projection measurement (ideal measurement) only, and thus 
the state of the system is projected onto the corresponding eigenspace of 
$H_{0}=H_{A}+H_{B}$ after the energy measurement. 
%ŽË‰e‰‰ŽZŽq'Ì'è‹`
Using the eigenstate $ |m\rangle $, we can write the projection operator 
onto the eigenspace of the Hamiltonian $H_{0}$ 
with the eigenvalue $E_{m} = E^{A}_{m_{A}} + E^{B}_{m_{B}} $ as 
%	ƒGƒlƒ‹ƒM[ŒÅ—Ló'Ԃւ̎ˉe‰‰ŽZŽq
\begin{align}
\Pi_{m} = |m\rangle \langle m|.
%\label{Projector}
\end{align}
%		Y
If the Hamiltonian $H_{0}$ has degeneracy, 
we distinguish the degenerate eigenstates with a quantum number $\lambda_m$ as 
$|m, \lambda_m \rangle$. 
In this case, the projection operator onto the eigenspace of $H_{0}$ 
with the eigenvalue $E_{m}$ is given by 
%	k'Þ'ª' 'éê‡'̎ˉe‰‰ŽZŽq
\begin{align}
\Pi_{m} = \sum_{\lambda_m} |m, \lambda_m\rangle \langle m, \lambda_m|.
%\label{Projector}
\end{align}
%		Y
In the following discussion, we assume the eigenvalues of $H_0$ to be 
nondegenerate for simplicity. 
Although Jarzynski and W\'{o}jcik did not 
mention the degeneracy of the Hamiltonians in their original paper,\cite{JarzynWo} 
the degeneracy indeed causes no corrections to their discussions.

%‰Šúó'Ԃ̍ì'è•û
For time $t<0$, each of the systems $A$ and $B$ is separately connected to a heat reservoir at the inverse temperatures $\beta_A$ and $\beta_B$, respectively, 
for sufficiently long time to reach its equilibrium state. 
The reservoirs are removed just before $t=0$, 
and hence the initial state of the total system is given by the following product state:
%‰Šúó'Ô'ªGibbsó'Ô
\begin{align}
\rho_{\rm init}=&\rho_{A} \otimes \rho_{B},
\label{Gibbssian}
\\
\rho_{\alpha}=&\frac{{\rm e}^{-\beta_\alpha H_{\alpha}}}{Z_{\alpha}},
\end{align}
where $Z_\alpha={\rm Tr}{\rm e}^{-\beta_\alpha H_{\alpha}}$
 is the partition function of the system $\alpha$ $(\alpha =A, B)$. 
%
%	Stage1
At $t=0$, we perform the first measurement of the energy of each system. 
Suppose that we obtained the outcome $(E^{A}_{m_A}, E^{B}_{m_B})$. 
Then, the state of the systems is projected onto the corresponding eigenstate of 
$H_0=H_A \otimes 1_B + 1_A \otimes H_B$: 
$|m\rangle=|m_A\rangle \otimes |m_B\rangle$ 
with the probability 
%	Å‰'Ì'ª'茋‰Ê'ª"¾'ç'ê'éŠm—¦
\begin{align}
\left\langle m \right|\rho_{\rm init}\left|m\right\rangle
%\left( |m\rangle, \rho_{\rm init}|m\rangle \right) 
	=
		\frac{e^{-\beta_A E^{A}_{m_A} -\beta_B E^{B}_{m_B}}}{Z_A Z_B}.
\end{align}
%		Y
%where $(\Phi , \Psi )$ express the inner product between arbitrary states $|\Psi \rangle$ 
%and $|\Phi \rangle$. 

The coupling between the two systems is turned on just after the first measurement 
at $t=0$. 
Then, the total system evolves according to the von Neumann equation from $t=0$ to $t=\tau$, 
where the time evolution is described by  
%ŽžŠÔ"­"W‰‰ŽZŽq
\begin{align}
U(t)={\rm e}^{-i(H_0+\gamma H_{\rm c})t/\hbar}.
\nonumber
\end{align}
At $t=\tau$, we separate the two systems again and measure the energy of each system. 
Suppose that we obtained the outcome $(E^{A}_{n_A}, E^{B}_{n_B})$. 
Then the state of the systems is projected onto the corresponding eigenstate of $H_0$: 
$|n\rangle=|n_A\rangle \otimes |n_B\rangle$.
The joint probability with which we observe $(E^{A}_{m_A},\ E^{B}_{m_B})$ at time $t=0$ 
and $(E^{A}_{n_A},\ E^{B}_{n_B})$ at time $t=\tau$ in the above process is given by 
\begin{align}
P(m , n | \tau) 
%	&=
%		\left( |m\rangle, \rho_{\rm init}|m\rangle \right) 
%		\ T_{m \to n}(\tau) 
%	\nonumber \\
	&= 
		\frac{e^{-\beta_A E^{A}_{m_A} -\beta_B E^{B}_{m_B}}}{Z_A Z_B}
		T_{m \to n}(\tau), 
\label{TwoTimeProb}
\end{align}
%		Y
where 
%	'JˆÚŠm—¦
\begin{align}
T_{m \to n}(\tau)=\left| \left\langle n\right| U(\tau) \left|m\right\rangle \right|^2
%T_{m \to n}(\tau)=\left| \left( |n\rangle, U(\tau) |m\rangle \right) \right|^2
\label{TransProb}
\end{align}
is the transition probability from $|m\rangle$ to $|n\rangle$. 
Using the projection operator $\Pi_{m}$, 
we can write the joint probability~\eqref{TwoTimeProb} as 
%ƒWƒ‡ƒCƒ"ƒgŠm—¦'ðŽË‰e‰‰ŽZŽq'ðŽg'Á'Ä
\begin{align}
P(m , n | \tau) 
	&=
		 {\rm Tr}\left( 
		\Pi_{n}U(\tau) \Pi_{m}\rho_{\rm init} \Pi_{m} U^{\dagger}(\tau)
		\right)
\nonumber \\
	&=
		 {\rm Tr}\left( 
		\Pi_{n}U(\tau) \rho_{\rm init} \Pi_{m} U^{\dagger}(\tau)
		\right). 
\label{TwoTimeProbTr}
\end{align}
%		Y
In the degenerate case, the joint probability is given 
by the same form as Eq.~\eqref{TwoTimeProbTr}.

%ŽžŠÔ"½"]'Ώ̐«'ɂ'¢'Ä
For derivation of the exchange fluctuation theorem, 
we hereafter assume the time-reversal invariance of the system. 
The time-reversal operation in quantum mechanics is described by an antilinear 
operator $\Theta$.\cite{TRQuant1} 
%
%'O'R'P'U
\if0
The antilinearity is represented as 
 %ŽžŠÔ"½"]‰‰ŽZŽq'ÌŽ'«Ž¿
\begin{align}
\Theta \left( \alpha_1|\Psi \rangle + \alpha_2 |\Phi \rangle \right) 
	&= 
		\alpha_1^{*}\Theta |\Psi \rangle + \alpha_2^{*} \Theta |\Phi \rangle,
\label{AntiLinear}
\end{align}
%		Y 
where $\alpha^{*}$ is complex conjugate of a scaler number $\alpha$. 
%XFT'ÍŽžŠÔ"½"]'Ώ̐«'Ì'à'ƂŐ¬'è—§'B
%Using the time-reversal operator $\Theta$, 
\fi
%'O'R'P'U
%
The time-reversal invariance of the system is then expressed as 
%ŽžŠÔ"½"]'Ώ̐«The assumption of the time-reversal invariance is described as follows:
\begin{align}
%\Theta H_{A} =  H_{A} \Theta, 
\left[ \Theta, H_{A} \right] &= 0, 
\hspace{0.5cm} 
%\Theta H_{B} =  H_{B} \Theta, 
\left[ \Theta, H_{B} \right] = 0, 
%\hspace{0.5cm}
\label{TimeReversal}
\\
%\Theta  H_{\rm c} =  H_{\rm c}  \Theta. 
\left[ \Theta, H_{\rm c} \right] &= 0. 
\label{TimeReversalC}
\end{align}
%		Y
The commutation relations in Eq.~\eqref{TimeReversal} are equivalent to 
the commutation relation between the projection operator $\Pi_m$ and 
the time-reversal operator $\Theta$: 
%ŽžŠÔ"½"]'Ώ̐«'ðŽË‰e‰‰ŽZŽq'Å•\Œ»
\begin{align}
\left[ \Theta, \Pi_{m} \right] = 0. 
\label{TimeReversalProj}
\end{align}
%		Y

By inserting the identity $1=\Theta^{-1} \Theta$ 
between $U(\tau)$ and $\rho_{\rm init}$ in Eq.~\eqref{TwoTimeProbTr}, 
and using Eqs.~\eqref{TimeReversal},~\eqref{TimeReversalC} 
and~\eqref{TimeReversalProj}, 
we have 
%	n'Æm'ð"ü'ê'Ö'¦'½'à'Ì
\begin{align}
P(m , n | \tau) 
	&=
		 {\rm Tr}\left( 
		\Pi_{n}U(\tau)\Theta^{-1} \Theta \rho_{\rm init} \Pi_{m} U^{\dagger}(\tau)
		\right)
\nonumber \\
%	&=
%		 {\rm Tr}\left( 
%		\Pi_{n}U(\tau)\Theta^{-1} \rho_{\rm init} \Pi_{m} \Theta U^{\dagger}(\tau)
%		\right)
%\nonumber \\
%	&=
%		 {\rm Tr}\left( 
%		\Pi_{n}U(\tau)\Theta^{-1} \rho_{\rm init} \Pi_{m} U(\tau) \Theta
%		\right)
%\nonumber \\
	&=
		 {\rm Tr}\left( 
		\Pi_{n}U^{\dagger}(\tau) \rho_{\rm init} \Pi_{m} U(\tau)
		\right) 
\nonumber \\
%	&=
%		\frac{e^{-\beta_A E^{A}_{m_A} -\beta_B E^{B}_{m_B}}}{Z_A Z_B}
%		{\rm Tr}\left( 
%		\Pi_{n}U^{\dagger}(\tau) \Pi_{m} U(\tau)
%		\right) 
%\nonumber \\
	&=
		\frac{e^{-\beta_A E^{A}_{m_A} -\beta_B E^{B}_{m_B}}}{Z_A Z_B}
		{\rm Tr}\left( 
		\Pi_{m} U(\tau)\Pi_{n}U^{\dagger}(\tau) 
		\right), 
\label{JointProbR}
\end{align}
%		Y
where we used $\Theta U(t)^{\dagger}=U(t)\Theta$. 
From Eq.~\eqref{JointProbR}, 
we obtain a symmetry relation about the joint probability 
%under the time-reversal invariance  
as follows: 
%ƒWƒ‡ƒCƒ"ƒgŠm—¦'ª–ž'½'·'Ώ̐«
\begin{align}	
\frac{P(m, n|\tau)}{P(n, m|\tau)}
	&=
		e^{-\beta_A \left(E^{A}_{m_A}-E^{A}_{n_A} \right) 
			-\beta_B \left(E^{B}_{m_B}-E^{B}_{n_B}\right)}
\nonumber \\
	&=
		{\rm e}^{\Delta \beta Q_{m \to n}} 
		{\rm e}^{\beta_B \Delta E_{m \to n}}, 
\label{JointProbSym}
\end{align}
%		Y
where 
%	"M'Ì'è‹`
\begin{align}
Q_{m \to n}=E^{A}_{m_A}-E^{A}_{n_A}
\end{align}
%		Y
is the energy decrease 
in the system $A$, 
%	'S'̂̃Gƒlƒ‹ƒM[•Ï‰»
\begin{align}
\Delta E_{m \to n}=E^{A}_{n_A}+E^{B}_{n_B}-E^{A}_{m_A}-E^{B}_{m_B}
\end{align}
%		Y 
is the energy change of the total system 
caused by the on and off of the coupling Hamiltonian 
and
%	‰·"x·
\begin{align} 
\Delta \beta = \beta_B - \beta_A. 
\end{align}
%		Y
We interpret 
%$Q^{A}_{m \to n}$ 
$Q_{m \to n}$ 
as the heat draining from the system $A$ 
to the system $B$. 
Jarzynski and W\'{o}jcik\cite{JarzynWo} have derived Eq.~\eqref{JointProbSym} 
in the same situation.

%'±'±'Ì•\Œ»'Í'¤'Ü'¢'Ì'ÅŽŸ'̏͂̃Cƒ"ƒgƒ'©'Ü'Æ'ß'É—p'¢'Ä'à'¢'¢'ÆŽv'¤B
\if0
When the coupling between the two systems is weak enough to 
guarantee the exchange fluctuation theorem, 
there is no heat transfer between the two systems 
and then the exchange fluctuation reduces to a trivial equality 
$p_{\tau}(Q)=\delta(Q)={\rm e}^{0}\delta(-Q)$.
Even though the exchange fluctuation theorem does not hold, 
there would be a possibility that the integral exchange fluctuation theorem 
Eq.~\eqref{IXFTOriginal} holds. However, we show that this integral fluctuation 
theorem does not hold either.
\fi

%
%	'±'̏ã'ÌŒ¾'¢‰ñ'µ'Ì'±'ƂˁB
%
 
%Verification of the XFT '͉¼Œ©o'µB
%
%'±'±'̃Cƒ"ƒgƒ'ÍŒã'ő啝'ɉü'è'·'é—\'è'P'Q'Q'U
\subsection{Deviation from the XFT and the IXFT}\label{Sec:Deviation}
Let us now retrace the derivation of the exchange fluctuation theorem\cite{JarzynWo}
and show the deviation from the theorem. 
Using the joint probability~\eqref{TwoTimeProb}, 
the probability distribution of the heat transfer $Q_{m \to n}$ is given by 
%	"M'Ì•ª•zŠÖ"'Ì'è‹`
\begin{align}
p_{\tau}(Q)=\sum_{m, n} P(m, n | \tau) \delta(Q-Q_{m \to n}), 
\label{DefProbQ}
\end{align}
where $\delta(\cdot)$ is the delta function. 
Using the symmetry relation of the joint probability~\eqref{JointProbSym} 
in Eq.~\eqref{DefProbQ}, 
we have  
%'JˆÚŠm—¦'ð'f'¼'É"WŠJ'µ'ÄXFT'Ì•\Ž®'É'ã"ü
\begin{align}
p_{\tau}(Q) - {\rm e}^{\Delta \beta Q} p_{\tau}(-Q)
	=
		{\rm e}^{\Delta \beta Q}
		\sum_{m , n}
		P(n, m | \tau)
		\left(
		{\rm e}^{\beta_B \Delta E_{m \to n} } -1 
		\right)
		 \delta\left(Q-Q_{m \to n}\right). 
\label{DeviationXFT}
\end{align}
%		Y

Jarzynski and W\'{o}jcik\cite{JarzynWo} assumed that 
the total energy of the systems $A$ and $B$ is almost preserved 
between the two measurements 
if the coupling $\gamma$ between the two systems is sufficiently weak; 
%'ª'茋‰Ê'̃Gƒlƒ‹ƒM['ª•Û'¶
\begin{align}
\Delta E_{m \to n}=E_{n} -E_{m} \simeq 0. 
\label{ApproxEne}
\end{align}
%		Y
In other words, they assumed 
\begin{align}
P(m, n|\tau) \propto \delta_{\Delta E_{m \to n}, 0}, 
\end{align}
%		Y
in the weak coupling limit. 
The right-hand side of Eq.~\eqref{DeviationXFT} vanishes under this assumption, 
and thereby they obtained the exchange fluctuation theorem: 
%	XFT by ƒWƒƒƒ‹ƒ`ƒ"ƒXƒL[
\begin{align}
p_{\tau}(Q) \simeq e^{\Delta \beta Q}p_{\tau}(-Q). 
\end{align}
%		Y

However, we must pay attention to the behavior of the net heat transfer $\langle Q \rangle_{\tau}$ in the weak coupling limit 
because we are interested in a non-equilibrium system, 
or in the presence of a finite heat transfer.  
From this point of view, 
we examine whether the exchange fluctuation theorem holds or not 
in the presence of a finite heat transfer %in the present paper 
for weak coupling. 

%XFT'Ì"j'ê'ðIXFT'ŏE'¤˜b
In general, the deviation term (the right-hand side of Eq.~\eqref{DeviationXFT}) 
has a finite value. 
We can check this by examining the deviation 
from the integral exchange fluctuation theorem, 
which is obtained by multiplying Eq.~\eqref{DeviationXFT} by ${\rm e}^{-\Delta \beta Q}$ 
and integrating it over the heat transfer $Q$ as 
%XFT'Ì"j'ê'ðIXFT'ŏE'¤
\begin{align}
		\langle {\rm e}^{-\Delta \beta Q}\rangle_{\tau}-1
	&=
		\sum_{m , n}
		P(n, m | \tau)
		\left(
		{\rm e}^{\beta_B \Delta E_{m \to n} } -1 
		\right), 
\label{DeviationIXFT0}
\end{align}
where we used $\int dQ~p_{\tau}(Q)=1$. 

In order to compare the magnitude of the deviation from the exchange fluctuation theorem 
and 
the net heat transfer for weak coupling, %in weak coupling limit, 
we use the interaction picture and see the $\gamma$ dependence of 
the joint probability. 
%the time-evolution operator. 
The time-evolution operator is written as 
%'ŠŒÝì—p•\Ž¦'ð—p'¢'ÄŽžŠÔ"­"W‰‰ŽZŽq'ð''­B
\begin{align}
U(t)= {\rm e}^{-\frac{i}{\hbar}H_{0} t} {\rm e}^{-i\gamma C(t)}, 
\label{TimeEvolutionInt}
\end{align}
%		Y
where $C(t)$ is a Hermitian operator defined as follows: 
\begin{align}
	iC(t)
		&=
		\frac{i}{\hbar} \int^{t}_{0} ds
		{\rm e}^{\frac{i}{\hbar}H_{0} s} H_{\rm c}{\rm e}^{-\frac{i}{\hbar}H_{0} s}
\nonumber \\
		&=
		\sum_{k=0}^{\infty}\frac{1}{(k+1)!}\left( \frac{i}{\hbar}t \right)^{k+1} 
		\left( \delta_{H_0} \right)^{k} H_{\rm c}. 
\label{HermitianC} 
\end{align}
%		Y
Here we used ${\rm e}^{A}B{\rm e}^{-A}=\sum_{k=0}^{\infty} (\delta_{A})^k B$
and 
$\delta_{A}$ denotes the inner-derivation operator
%ŒðŠ·ŠÖŒW'ð'±'ê'Å•\'·B
 \begin{align}
 \delta_A = \left[ A, \ \right]. 
 \end{align}
%		Y 
Using Eq.~\eqref{TimeEvolutionInt} in the transition probability~\eqref{TransProb},  
we have 
%'JˆÚŠm—¦'𑊌ݍì—p•\Ž¦'ŏ''­
\begin{align}
T_{m\to n}(\tau) 
	&=
		\left| \langle n| {\rm e}^{-i\gamma C(\tau)} |m\rangle \right|^2
\nonumber \\
	&=
		{\rm Tr} \left( \Pi_n  {\rm e}^{-i\gamma C(\tau)} \Pi_m {\rm e}^{i\gamma C(\tau)} \right), 
\end{align}
%		Y
and the joint probability becomes 
%'JˆÚŠm—¦'𑊌ݍì—p•\Ž¦'ŏ''­
\begin{align}
P_{m\to n}(\tau) 
	&=
		{\rm Tr} 
		\left( \Pi_n  {\rm e}^{-i\gamma C(\tau)} 
		 \rho_{\rm init} \Pi_m {\rm e}^{i\gamma C(\tau)} \right).
\label{JointProbTr} 
\end{align}
%		Y
Using the joint probability in the trace form~\eqref{JointProbTr}, 
we have the deviation term from the integral exchange fluctuation theorem as 
%XFT'Ì"j'ê'ðIXFT'ŏE'¤
\begin{align}
		\langle {\rm e}^{-\Delta \beta Q}\rangle_{\tau}-1
	=&
		\sum_{m , n}
		 {\rm Tr}\left( 
		\Pi_{m}{\rm e}^{-i\gamma C(\tau)} \rho_{\rm init} \Pi_{n} 
		{\rm e}^{i\gamma C(\tau)}
		\right)
		\left(
		{\rm e}^{\beta_B \Delta E_{m \to n} } -1 
		\right) 
	\nonumber \\
	=&
		\sum_{m , n}
		{\rm Tr}\left(
		{\rm e}^{-\beta_B H_0 } 
		\Pi_m {\rm e}^{i\gamma C(\tau)} 
		\rho_{\rm init}
		{\rm e}^{\beta_B  H_0 } \Pi_n 
		{\rm e}^{-i\gamma C(\tau)}
		\right)
		\nonumber \\
	& 	-
		\sum_{m , n}
		{\rm Tr}\left(
		\Pi_m {\rm e}^{i\gamma C(\tau)} 
		\rho_{\rm init}\Pi_n 
		{\rm e}^{-i\gamma C(\tau)}
		\right)
		\nonumber \\
	=&
		{\rm Tr}\left(
		\rho_{\rm init}
		{\rm e}^{\beta_B  H_0 } 
		{\rm e}^{-i\gamma C(\tau)}
		{\rm e}^{-\beta_B H_0 } 
		{\rm e}^{i\gamma C(\tau)} 
		\right)
%		\nonumber \\
%	&\ \ 	
		-
		1.
%	\nonumber 
\end{align}
%		Y
We then expand the exponential ${\rm e}^{\pm i\gamma C(\tau)}$ to obtain 
\begin{align}
\langle {\rm e}^{-\Delta \beta Q}\rangle_{\tau}-1
	=&
		\sum_{k=0}^{\infty} 
		\frac{1}{k!}(-i\gamma)^k
		{\rm Tr}\left[
		\rho_{\rm init}
		{\rm e}^{\beta_B  H_0 } 
		 (\delta_{C(\tau)})^k {\rm e}^{-\beta_B H_0 }
		\right]
		-
		1
	\nonumber \\
	=&
		-i\gamma
		{\rm Tr}\left(
		\rho_{\rm init}
		{\rm e}^{\beta_B  H_0 } 
		 \left[C(\tau), {\rm e}^{-\beta_B H_0 }\right]
		\right) 
	\nonumber \\
	&	+ 
		\sum_{k=2}^{\infty} 
		\frac{1}{k!}(-i\gamma)^k
		{\rm Tr}\left[
		\rho_{\rm init}
		{\rm e}^{\beta_B  H_0 } 
		 (\delta_{C(\tau)})^k {\rm e}^{-\beta_B H_0 }
		\right]
	\nonumber \\
	=&
		-i\gamma
		{\rm Tr}\left(
		 \left[{\rm e}^{-\beta_B H_0 }, \rho_{\rm init}{\rm e}^{\beta_B  H_0 }\right]
		 C(\tau)
		\right)
	\nonumber \\
	&	+
		\sum_{k=2}^{\infty} 
		\frac{1}{k!}(-i\gamma)^k
		{\rm Tr}\left[
		\rho_{\rm init}
		{\rm e}^{\beta_B  H_0 } 
		 (\delta_{C(\tau)})^k {\rm e}^{-\beta_B H_0 }
		\right]
	\nonumber \\
	=&
		\sum_{k=2}^{\infty} 
		\frac{1}{k!}(-i\gamma)^k
		{\rm Tr}\left[
		\rho_{\rm init}
		{\rm e}^{\beta_B  H_0 } 
		 (\delta_{C(\tau)})^k {\rm e}^{-\beta_B H_0 }
		\right],
\label{DeviationIXFT}
\end{align}
%		Y
where we used $\left[{\rm e}^{\beta_B H_0}, \rho_{\rm init}\right]=0$ in the last line. 
This commutability is a consequence of taking the initial state as a Gibbs state. 
We thus see that the deviation is of the second order of $\gamma$ 
in the lowest order. 
In \S\ref{sec:Example}, the deviation is explicitly calculated 
for simple models and is shown to have a finite value for finite $\gamma$.

Next, 
we derive the net heat transfer in the power series of $\gamma$ and 
show that its lowest order of $\gamma$ is also the second order: 
%"M—¬'̃Aƒ"ƒTƒ"ƒuƒ‹•½‹Ï'ðƒgƒŒ[ƒX'©'Â\gamma'ׂ̂«'ŏ''­
\begin{align}
	\langle Q \rangle_{\tau}
	=&
		\int^{\infty}_{\infty} d Q\ p_{\tau}(Q)\ Q
	\nonumber \\
	=&
		\sum_{m, n} 
		{\rm Tr}
		\left(
		\Pi_n {\rm e}^{-i\gamma C(\tau)} 
		\rho_{\rm init} \Pi_m 
		{\rm e}^{i\gamma C(\tau)}
		\right)
		\left( E^{A}_{m_A}-E^{A}_{n_A} \right)
	\nonumber \\
	=&
		\sum_{m, n} 
		{\rm Tr}
		\left(
		\Pi_n {\rm e}^{-i\gamma C(\tau)} 
		\rho_{\rm init} 
		H_A
		\Pi_m 
		{\rm e}^{i\gamma C(\tau)}
		-
		H_A
		\Pi_n {\rm e}^{-i\gamma C(\tau)} 
		\rho_{\rm init} 
		\Pi_m 
		{\rm e}^{i\gamma C(\tau)}
		\right)
	\nonumber \\
	=&
		{\rm Tr}
		\left(
		{\rm e}^{-i\gamma C(\tau)} 
		\rho_{\rm init} 
		H_A 
		{\rm e}^{i\gamma C(\tau)}
		-
		H_A
		{\rm e}^{-i\gamma C(\tau)} 
		\rho_{\rm init} 
		{\rm e}^{i\gamma C(\tau)}
		\right)
	\nonumber \\
	=&
		{\rm Tr}
		\left(
		\rho_{\rm init} 
		H_A 
		-
		\rho_{\rm init} 
		{\rm e}^{i\gamma C(\tau)}
		H_A
		{\rm e}^{-i\gamma C(\tau)} 
		\right).
%	\nonumber 
\end{align}
%		Y
We again expand the exponential ${\rm e}^{\pm i \gamma C(\tau)}$ to obtain 
\begin{align}
\langle Q \rangle_{\tau}
	=&
		{\rm Tr}
		\rho_{\rm init} 
		H_A 
		-
		\sum_{k=0}^{\infty}
		\frac{1}{k!}(i\gamma)^k
		{\rm Tr}
		\left[
		\rho_{\rm init} 
		(\delta_{C(\tau)})^k
		H_A
		\right]
	\nonumber \\
	=&
		-i\gamma
		{\rm Tr}
		\left(
		\rho_{\rm init} 
		[C(\tau), H_A]
		\right)
		-
		\sum_{k=2}^{\infty}
		\frac{1}{k!}(i\gamma)^k
		{\rm Tr}
		\left[
		\rho_{\rm init} 
		(\delta_{C(\tau)})^k
		H_A
		\right]
	\nonumber \\
	=&
		-i\gamma
		{\rm Tr}
		\left(
		C(\tau) 
		[H_A, \rho_{\rm init}]
		\right)
		-
		\sum_{k=2}^{\infty}
		\frac{1}{k!}(i\gamma)^k
		{\rm Tr}
		\left[
		\rho_{\rm init} 
		(\delta_{C(\tau)})^k
		H_A
		\right]
	\nonumber \\
	=&
		-
		\sum_{k=2}^{\infty}
		\frac{1}{k!}(i\gamma)^k
		{\rm Tr}
		\left[
		\rho_{\rm init} 
		(\delta_{C(\tau)})^k
		H_A
		\right],
\label{EQ}
\end{align}
%		Y
where we used $\left[\rho_{\rm init}, H_A\right]=0$ in the last line. 
The lowest order of $\gamma$ in $\langle Q \rangle_{\tau}$ is the second order, 
which is the same as the deviation term in Eq.~\eqref{DeviationIXFT}. 
This means that the smaller the deviation term becomes for weak coupling, 
the less the net heat transfer between the two systems becomes. 
We can show that the higher moments of $p_{\tau}(Q)$ has the same dependence of 
$\gamma$, and thus the exchange fluctuation theorem becomes just a trivial relation 
in the limit $\gamma \to 0$: 
$p_{\tau}(Q)=\delta(Q)={\rm e}^{\Delta \beta Q}\delta(-Q)=\delta(-Q)$. 

In \S\ref{sec:Example}, 
we show for specific models 
that 
the deviation term 
$\langle {\rm e}^{-\Delta \beta Q}\rangle_{\tau}-1$ and 
the net heat transfer $\langle Q \rangle_{\tau}$ 
are, in general, both finite in the second order of $\gamma$.

%ƒRƒ"ƒXƒ^ƒ"ƒgƒJƒbƒvƒŠƒ"ƒO'Å'Q''̌n'ðŒq'°'é'ÆXFT'ªŒµ–§'È"™Ž®'Æ'µ'Ä
%–ž'½'³'ê'邱'Æ'ðŒ©'éB
\section{Commutable-coupling condition} \label{sec:CCC}
We showed in the previous section that the exchange fluctuation theorem 
has a finite deviation term in general, as far as the coupling strength $\gamma$ is finite. 
Before confirming this for specific models, 
we present an additional condition for which the exchange 
fluctuation theorem and the integral exchange fluctuation theorem hold exactly 
under a finite heat transfer and for arbitrary coupling strength $\gamma$. 
Our additional condition is 
%CCC
\begin{align}
\left[H_0, H_{\rm c}\right]=0, 
\label{CCC}
\end{align}
%		Y
which we refer to as the commutable-coupling condition. 
Jarzynski's equality for open quantum systems was discussed under this condition.\cite{CCCpaper} 
With the commutable-coupling condition, 
the energy of the system 
$H_0$ 
is conserved throughout the process 
and hence $\Delta E_{m \to n}=0$ holds exactly in Eq.~\eqref{DeviationXFT}. 

Let us be more precise. Under the commutable-coupling condition, 
we can separate the Hamiltonian $H_0$ and $H_{\rm c}$ in the time-evolution operator, 
and thus the transition probability~\eqref{TransProb} is reduced to  
%'JˆÚŠm—¦'Å H_0'ÆH_c '𕪗£
\begin{align}
T_{m \to n}(\tau)
	&=
		|\langle n | {\rm e}^{-i\gamma\frac{H_{\rm c}}{\hbar}\tau } | m \rangle|^2. 
\end{align}
%		Y
Using the commutable-coupling condition~\eqref{CCC} again, 
we have
\begin{align}
\left( E_{n}-E_{m} \right)\langle n |{\rm e}^{-i\gamma\frac{H_{\rm c}}{\hbar}\tau } | m \rangle
	= 
	\langle n | H_{0} {\rm e}^{-i\gamma\frac{H_{\rm c}}{\hbar}\tau } | m \rangle
	-
	\langle n |{\rm e}^{-i\gamma\frac{H_{\rm c}}{\hbar}\tau } H_{0}| m \rangle
	=
	0, 
%\label{EnePreserveCCC}
\end{align}
%		Y
and therefore we obtain 
\begin{align}
\langle n |{\rm e}^{-i\gamma\frac{H_{\rm c}}{\hbar}\tau } | m \rangle
	= 
		\langle n |{\rm e}^{-i\gamma\frac{H_{\rm c}}{\hbar}\tau } | m \rangle
		\delta_{\Delta E_{m \to n}, 0}. 
%		\delta_{E_{n}, E_{m}}. 
\label{EnePreserveCCC}
\end{align}
%		Y
We can see that 
the energy of the total system is preserved 
throughout the two-point measurement process. 
Using Eq.~\eqref{EnePreserveCCC}, we have 
\begin{align}
P(m, n|\tau)\propto T_{m\to n}(\tau) \propto \delta_{\Delta E_{m\to n}, 0}, 
\end{align}
%		Y
and thereby conclude that the exchange fluctuation theorem becomes an exact relation 
under the commutable-coupling condition:
%CCC‰º'ÅXFT'ª¬—§
\begin{align}
&p_{\tau}(Q) - {\rm e}^{\Delta \beta Q} p_{\tau}(-Q)
	\nonumber \\
	&=
		{\rm e}^{\Delta \beta Q}
		\sum_{m , n}
		P(n, m | \tau)
		\delta_{\Delta E_{m \to n}, 0} 
%		\delta_{E_{n}, E_{m}}. 
		\left(
		{\rm e}^{\beta_B \Delta E_{m \to n} } -1 
		\right)
		 \delta\left(Q-Q_{m \to n}\right) 
	=0.
\end{align}
%		Y
The integral exchange fluctuation theorem also holds exactly 
under the commutable-coupling condition, 
since it is a direct consequence of 
the exchange fluctuation theorem. 

Next, we show that a finite heat transfer between the two systems does exist under 
the commutable-coupling condition. 
Under the commutable-coupling condition, the Hermitian operator $C(t)$
becomes 
%CCC‰º'Å'ÌC(t)
\begin{align}
C(t)
		=\frac{1}{\hbar} \int^{t}_{0} ds
		{\rm e}^{\frac{i}{\hbar}H_{0} s} H_{\rm c}{\rm e}^{-\frac{i}{\hbar}H_{0} s}
		=\frac{t}{\hbar}H_{\rm c}. 
\label{HermitianCwithCCC}
\end{align}
%		Y
From Eqs.~\eqref{EQ} and~\eqref{HermitianCwithCCC}, 
the ensemble average of the heat transfer can be  written in the trace form: 
%	CCC‰º'Å'Ì"M'ÌŠú'Ò'l
\begin{align}
	\langle Q \rangle_{\tau}
	&=		
		-\sum_{k=2}^{\infty}
		\frac{1}{k!} \left(i \frac{\gamma \tau}{\hbar}\right)^k
		{\rm Tr}
		\left[
			\rho_{\rm init}
			\left( \delta_{H_{\rm c}} \right)^{k} H_A
		\right].
\nonumber
%\label{EQCCC}
\end{align}
%		Y
Note that 
 $\delta_{H_{\rm c}} H_{A}$ 
is generally finite, 
%in the commutable-coupling condition, 
although we assume $\delta_{H_{\rm c}} H_{0}=\delta_{H_{\rm c}} (H_{A}+H_{B})=0$.  
Therefore the heat can flow between the two systems
and 
the exchange fluctuation theorem is a nontrivial relation 
under the commutable-coupling condition. 
We check this for specific models in the next section. 

%%%%%
\section{Examples}\label{sec:Example}
\subsection{A two-spin 1/2 system}\label{Sec:Two-Spin}
\begin{figure}
\includegraphics[width=\textwidth]{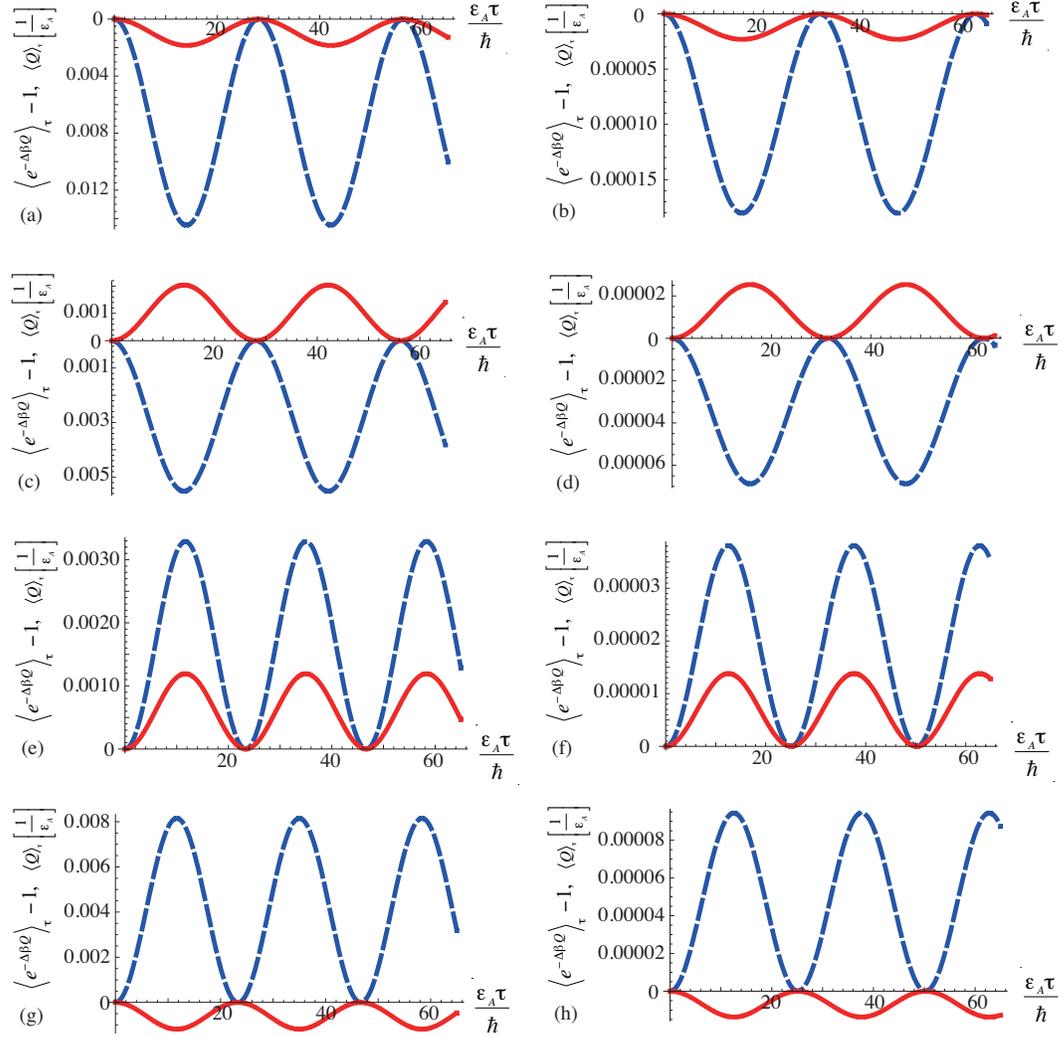}
	\caption	{
		The quantities $\langle {\rm e}^{-\Delta \beta Q} \rangle_{\tau}-1$ 
		(solid red line) 
		and $\langle Q\rangle_{\tau}$ (broken blue line) in the two-spin 1/2 system. 
		We fixed $\beta_A \epsilon_A=2$: 
		for 
		$\epsilon_B / \epsilon_A	=0.8$, 
		$\beta_B / \beta_A		=0.9$ 
		with 
		(a) 
		$\gamma J / \epsilon_A			=0.1$ 
		and 
		(b) 
		$\gamma J / \epsilon_A			=0.01$;
		for  
		$\epsilon_B / \epsilon_A	=0.8$, 
		$\beta_B / \beta_A		=1.1$ 
		with 
		(c) 
		$\gamma J / \epsilon_A			=0.1$ 
		and
		(d) 
		$\gamma J / \epsilon_A			=0.01$; 
		for  
		$\epsilon_B / \epsilon_A	=1.25$, 
		$\beta_B / \beta_A		=0.9$ 
		with 
		(e) 
		$\gamma J / \epsilon_A			=0.1$ 
		and 
		(f) 
		$\gamma J / \epsilon_A			=0.01$; 
		for  
		$\epsilon_B / \epsilon_A	=1.25$, 
		$\beta_B / \beta_A		=1.1$ 
		with 
		(g) 
		$\gamma J / \epsilon_A			=0.1$ 
		and 
		(h) 
		$\gamma J / \epsilon_A			=0.01$. 
		}
		\label{Fig:TwoSpin}
\end{figure}

As the first example, 
we consider a quantum system which consists of two spin $1/2$s 
initially prepared at different temperatures. 
The spins exchange the energy via the Heisenberg coupling between the two spins. 
Thus, this system is given by the following Hamiltonian:
%Hamiltonian'Ì'è‹`
\begin{align}
H &= H_0 + \gamma H_{\rm c},  
\nonumber \\ 
H_0 &=H_A \otimes 1_{B} + 1_A \otimes H_B,
\nonumber 
\end{align}
%		Y
where $H_{A}$, $H_{B}$ and $H_{\rm c}$ are 
\begin{align}
H_{A} &= - \frac{\epsilon_{A}}{2} \sigma_{A}^z, 
%&H_{A} = - \frac{\epsilon_{A}}{2} \sigma_{A}^z
%		, \ \ \ \ \ \epsilon_A = \hbar \Omega_A, 
		\\
H_{B} &= - \frac{\epsilon_{B}}{2} \sigma_{B}^z, 
%&H_{B} = - \frac{\epsilon_{B}}{2} \sigma_{B}^z 
%		,\ \ \ \ \ \epsilon_B = \hbar \Omega_B, 
		\\
H_{\rm c}&= -\frac{J}{4} \vec{\sigma}_{A} \cdot \vec{\sigma}_{B}
%&H_{\rm c} = -\frac{J}{4} \vec{\sigma}_{A} \cdot \vec{\sigma}_{B}.
%		 \ \  J=\hbar^{2} j, 
\end{align}
with $ \sigma^x,  \sigma^y$ and $ \sigma^z$ denoting the Pauli matrices. 
%'O'X@'O'Q'P'T
\if0
%ƒpƒEƒŠs—ñ
\begin{align}
\sigma^{x}=
\begin{pmatrix}
0 & 1 \\
1 & 0 
\end{pmatrix}
, 
\hspace{1.0cm}
\sigma^{y}=
\begin{pmatrix}
0 & -i \\
i & 0 
\end{pmatrix}
, 
\hspace{1.0cm}
\sigma^{z}=
\begin{pmatrix}
1 & 0 \\
0 & -1
\end{pmatrix}.
\end{align}
%
\fi
%'O'X@'O'Q'P'T
This model is analytically solvable and the deviation 
from the integral exchange fluctuation theorem, 
$\langle {\rm e}^{-\Delta \beta Q}\rangle_{\tau}-1$,
and the heat transfer $\langle Q \rangle_{\tau}$ are calculated explicitly as follows:
%'QƒXƒsƒ"'Ì—á'Å‹ï'Ì"I'Ȋ֐"Œ`'ðŽ¦'·B
%XFT'̐ϕªŒ^
\begin{align}
\langle {\rm e}^{-\Delta \beta Q} \rangle_{\tau}-1
	=& 		\frac{\gamma^2 J^2}
			{(\epsilon_{A}-\epsilon_{B})^2+\gamma^2 J^2}
		\mathop{\mathrm{sech}} 
		\left[ 
			\frac{\beta_A \epsilon_A}{2}
		\right]
		\mathop{\mathrm{sech}} 
		\left[ 
			\frac{\beta_B \epsilon_B}{2}
		\right] 
		\nonumber \\
	&	\times 
		\sinh 
		\left[
			\frac{\epsilon_A}{2}
			\left( 
				\beta_B - \beta_A
			\right)
		\right]
		\sinh 
		\left[
			\frac{\beta_B}{2}
			\left( 
			\epsilon_A-\epsilon_B
			\right)
		\right]
		\nonumber \\
	& 	\times 
		\sin^2
		\left(
			\frac{\tau}{2\hbar}
			\sqrt{(\epsilon_{A}-\epsilon_{B})^2+\gamma^2 J^2}
		\right),
\label{DeviationIXFTTwoSpin}
%\end{align}
%		Y
%
\\
%"M—¬'ÌŠú'Ò'l
%\begin{align}
\langle Q \rangle_{\tau}
%	&=
%		\epsilon_{A} 
%		\frac{{\rm e}^{\beta_{B} \epsilon_{B}}-{\rm e}^{\beta_{A} \epsilon_{A}}}
%			{(1+{\rm e}^{\beta_{A} \epsilon_{A}})(1+{\rm e}^{\beta_{B} \epsilon_{B}})}
%		\frac{J^2}
%			{(\epsilon_{A}-\epsilon_{B})^2+J^2}
%		\nonumber \\
%	&\ \ 	\times 
%		\sin^2
%		\left[
%			\frac{\tau}{4\hbar}
%			\left(
%				\epsilon_{A}+\epsilon_{B}-J+\sqrt{(\epsilon_{A}-\epsilon_{B})^2+J^2}
%			\right)
%		\right].
	%'à'µ'­'Í
%	\\
	=&
		\frac{\epsilon_{A}}{2} 
		\frac{\gamma^2 J^2}
			{(\epsilon_{A}-\epsilon_{B})^2+\gamma^2 J^2}
		\mathop{\mathrm{sech}} 
		\left[ 
			\frac{\beta_A \epsilon_A}{2}
		\right]
		\mathop{\mathrm{sech}} 
		\left[ 
			\frac{\beta_B \epsilon_B}{2}
		\right] 
		\nonumber \\
	& 	\times 
		\sinh 
		\left[ 
			\frac{1}{2}
			\left( 
				\beta_B \epsilon_B - \beta_A \epsilon_A
			\right)
		\right]		
		\nonumber \\
	& 	\times 
		\sin^2
		\left(
			\frac{\tau}{2\hbar}
			\sqrt{(\epsilon_{A}-\epsilon_{B})^2+\gamma^2 J^2}
		\right).
\label{EQTwoSpin}
\end{align}
%		Y
The $\tau$ dependences of the deviation term $\langle {\rm e}^{-\Delta \beta Q} \rangle_{\tau}-1$ 
and 
the heat transfer $\langle Q \rangle_{\tau}$ are 
shown in Fig.~\ref{Fig:TwoSpin}.
%%%%%%%%%%%%%%%%%%
%%%%%%%%%%%%%%
%%%%%%%%%%%
%%%%%%%%%%%		TwoSpin'ÌXFTƒYƒŒ'Æ"M—¬
%		ƒpƒ‰ƒƒ^'̏‡"Ô		\epsilon_B,  \beta_A,  \gamma \beta=\beta_B//beta_A
%%%
%'QƒXƒsƒ"'̃Oƒ‰ƒt
%%
%\epsilon_B = 1.25, \beta_A=1, \beta_B=1.1 'Ì"äŠrI'í'è
%%%%%%%%%%%		TwoSpin'ÌXFTƒYƒŒ'Æ"M—¬
%%%%%%%%%%%%%%

\begin{figure}%[htbp]
\centering
\includegraphics[width=0.7\textwidth]{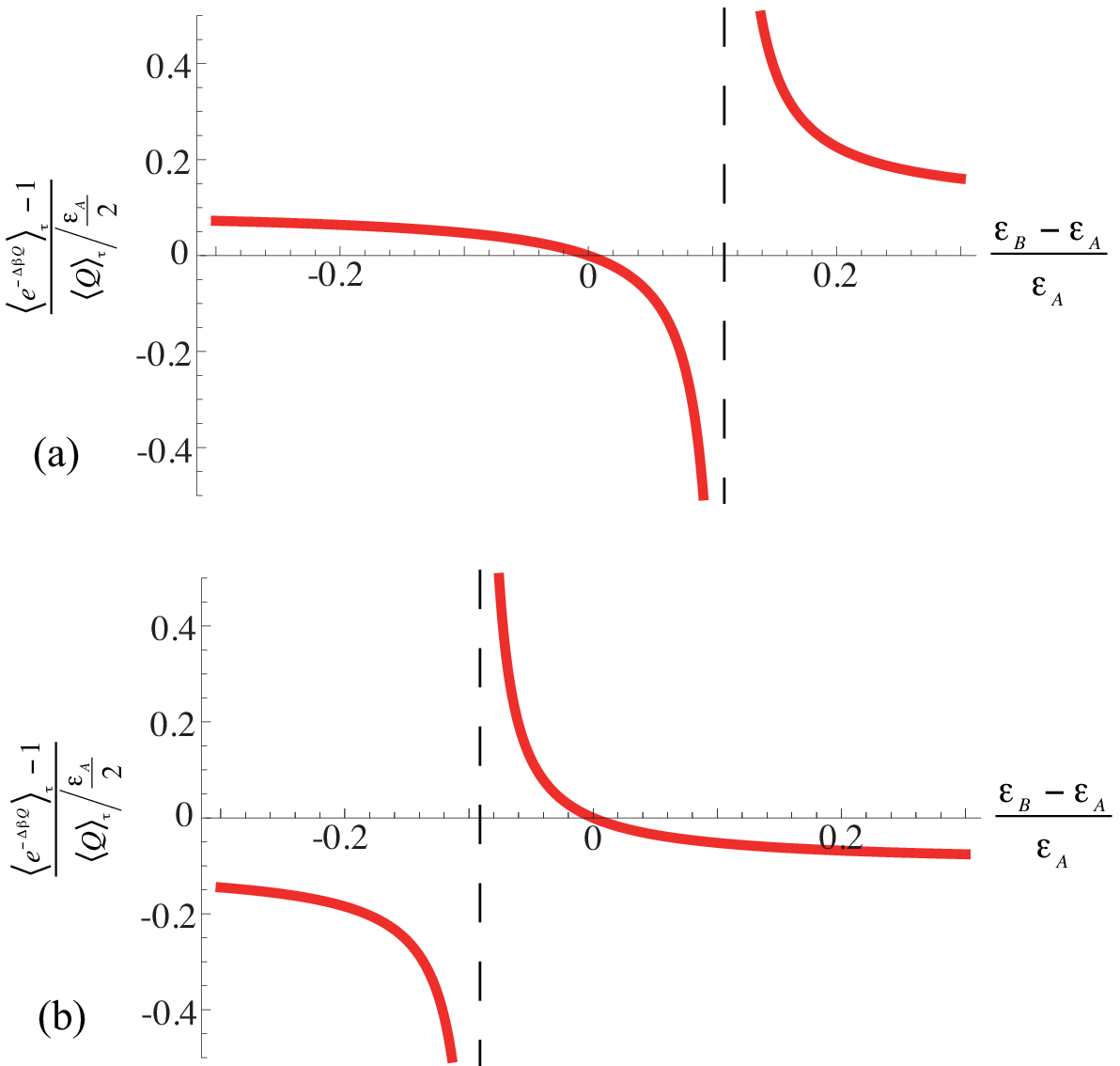}
%	\begin{center}
%		\includegraphics[width=10cm]{TwoSpinRatioFig_aVer3.eps}
%	\end{center}
%%\label{TwoSpinRatio1_11}
%%\end{figure}
%%
%%
%%		'Q'–Ú
%%\begin{figure}%[htbp]
%	\begin{center}
%		\includegraphics[width=10cm]{TwoSpinRatioFig_bVer3.eps}
%	\end{center}
	\caption{The ratio of the 
	$\langle {\rm e}^{-\Delta \beta Q} \rangle_{\tau}-1$ to $\langle Q \rangle_{\tau}$. 
	We fixed   $\beta_A \epsilon_A=2$ with  
	(a) 
	$\beta_B / \beta_A=0.9$ 
	and 
	(b)
	$\beta_B / \beta_A=1.1$. 
	The commutable-coupling condition is fulfilled at the point $\epsilon_B - \epsilon_A=0$.} 
\label{Fig:TwoSpinRatio}
\end{figure}

Note that $\langle {\rm e}^{-\Delta \beta Q} \rangle_{\tau}-1$ 
and $\langle Q \rangle_{\tau}$ have the same dependence on $\gamma$ in this model. 
Thus, the ratio of these two quantities is independent of 
the coupling strength $\gamma$: 
%XFT'Æ"M—¬'Ì"ä
\begin{align}
\frac{\langle {\rm e}^{-\Delta \beta Q} \rangle_{\tau}-1}
	{\langle Q \rangle_{\tau}/(\frac{\epsilon_A}{2})}
		&=
		\frac
		{\sinh \left[ \frac{\epsilon_A}{2}(\beta_B-\beta_A) \right]
		\sinh \left[ \frac{\beta_B}{2}(\epsilon_A-\epsilon_B) \right]
		}		
		{
		\sinh \left[ \frac{1}{2}(\beta_B \epsilon_B - \beta_A \epsilon_A) \right]
		}.
%	\nonumber \\
%		&=
%		\frac
%		{\sinh \left[ \frac{\epsilon_A}{2} \Delta \beta \right]
%		\sinh \left[ \frac{\Delta \beta+\beta_A}{2}(\epsilon_A-\epsilon_B) \right]
%		}		
%		{
%	\sinh \left\{
%		\frac{1}{2}\left[ (\Delta \beta \epsilon_B+\beta_A(\epsilon_B-\epsilon_A) \right]
%		\right\}
%		}. 
\label{RatioTwoSpin}
\end{align}
%		Y
Figure~\ref{Fig:TwoSpinRatio} shows the parameter dependence of the ratio. 
The ratio has a finite value for almost all range of the energy-level difference 
$\epsilon_B - \epsilon_A$. 
%%%%%%%%%%
%%%%%%		XFT'©'ç'̃YƒŒ'Æ"M—¬'ÌŠú'Ò'l'Ì"ä'ðƒvƒƒbƒg
%%%	
%%%	
%%%%%%		XFT'©'ç'̃YƒŒ'Æ"M—¬'ÌŠú'Ò'l'Ì"ä'ðƒvƒƒbƒg'µ'½
%%%%%%%%%%
As a consequence, if the coupling strength $\gamma$ is weak enough to neglect 
the deviation from the integral exchange fluctuation theorem, 
Eq.~\eqref{DeviationIXFTTwoSpin}, 
the net heat transfer $\langle Q \rangle_{\tau}$ in Eq.~\eqref{EQTwoSpin} 
is also negligibly small. 
This result clearly shows that the exchange fluctuation theorem does not generally hold 
in the presence of a finite heat transfer. 
%and the rang of the validity of the exchange fluctuation theorem 
%is restricted to near-equilibrium states.

%Note, however, that the ratio in Fig.~\ref{Fig:TwoSpinRatio} 
However, the ratio in Fig.~\ref{Fig:TwoSpinRatio} 
vanishes for $\epsilon_A = \epsilon_B$; that is, 
the integral exchange fluctuation theorem is 
recovered at this particular point with a finite heat transfer. 
At this point $\epsilon_A = \epsilon_B$ and at this point only, 
the commutable-coupling condition is satisfied: 
%	CCC in Two-Spin
\begin{align}
[H_0, H_{\rm c}]
	&=
		i\frac{J}{4}
		\left(
		\epsilon_A - \epsilon_B 
		\right)
		\left(
		\sigma^{y}_{A}\sigma^{x}_{B}-\sigma^{x}_{A}\sigma^{y}_{B}
		\right)
	=0.
\label{TwoSpinCCC}
\end{align}
%		Y

Under the commutable-coupling condition $\epsilon_A = \epsilon_B$, 
Eqs.~\eqref{DeviationIXFTTwoSpin} and \eqref{EQTwoSpin} become 
\begin{align}
&\langle {\rm e}^{-\Delta \beta Q} \rangle_{\tau}-1
	=0,
\\
&\langle Q \rangle_{\tau}
	=
		\frac{\epsilon}{2} 
		\mathop{\mathrm{sech}} 
		\left[ 
			\frac{\epsilon}{2}\beta_A 
		\right]
		\mathop{\mathrm{sech}} 
		\left[ 
			\frac{ \epsilon}{2}\beta_B
		\right] 
		\sinh 
		\left[ 
			\frac{\epsilon}{2}
			\left( 
				\beta_B - \beta_A 
			\right)
		\right] 
		\sin^2 
		\left[ \frac{\gamma J }{2\hbar} \tau \right].
\end{align}
%		Y
The net heat transfer has a finite value, 
while the integral exchange fluctuation theorem holds. %for any $\gamma$.
Thus we demonstrated in this model that 
the integral exchange fluctuation theorem holds 
if and only if the commutable-coupling condition is satisfied.

%			'Q'–ڂ̗á
%
%'²˜aU"®Žq'Ì—áB1223"úAJaynes-Cummings model '֕ύXB
\subsection{Coupled harmonic oscillators}
The second example is a system which consists of two harmonic oscillators.
This system is given by the following Hamiltonian:
%ƒnƒ~ƒ‹ƒgƒjƒAƒ"
\begin{align}
H &= H_0  + \gamma H_{\rm c}, 
\label{HamiltonianHO} \\
H_0 &= H_A \otimes 1_{B} + 1_A \otimes H_B. 
\end{align}
%		Y
Here the Hamiltonian $H_A$, $H_B$ and $H_{\rm c}$ are 
%ƒnƒ~ƒ‹ƒgƒjƒAƒ"'ðŒÂ•Ê'ÉŒ©'é'Æ
\begin{align}
H_{A}& = \hbar \omega_A a^{\dagger} a,				 \nonumber \\
H_{B} &= \hbar \omega_B b^{\dagger} b,				 \nonumber \\
H_{\rm c}& =\nu \left(  a^{\dagger} b  +  b^{\dagger} a \right),	 
\label{HOcoupling}
\end{align}
%		Y
where $a$ and $a^{\dagger}$ are the creation and annihilation operators 
of the oscillator $A$, $b$ and $b^{\dagger}$ are those of $B$, 
$\left[a, a^{\dagger} \right]=1, \left[a, a\right]=0$ 
and 
$\left[b, b^{\dagger} \right]=1, \left[b, b\right]=0$, 
and $\nu$ is a real number. 

\begin{figure}[t]
\includegraphics[width=\textwidth]{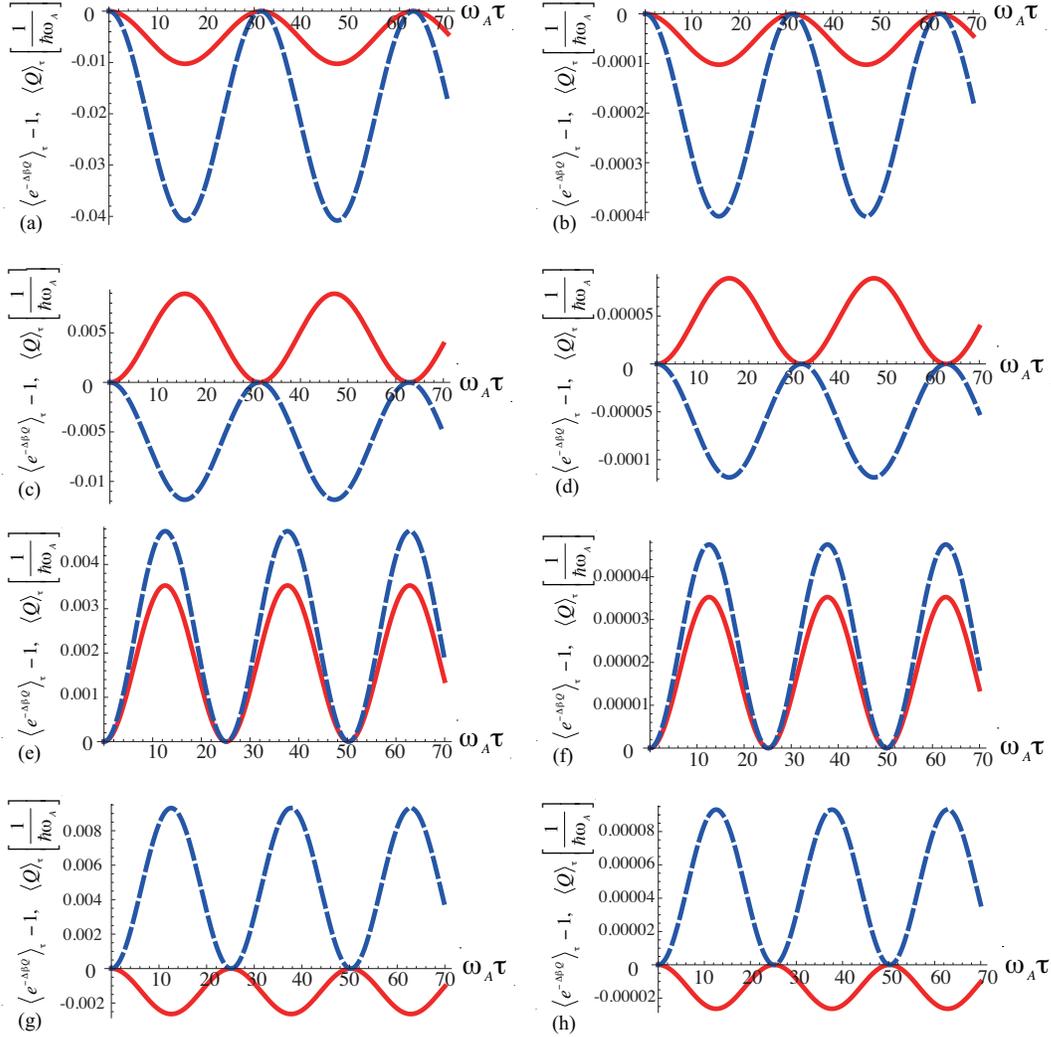}
%%\omega_B=0.8, \beta_A=4, \beta_B/\beta_A=0.9 'Ì"äŠr
%\begin{minipage}{0.5\hsize}
%\begin{center}
%\includegraphics[width=70mm]{HOg01_08_4_09.eps}
%\end{center}
%\end{minipage}
%\begin{minipage}{0.5\hsize}
%\begin{center}
%\includegraphics[width=70mm]{HOg001_08_4_09.eps}
%\end{center}
%\end{minipage}
%%\omega_B=0.8, \beta_A=4, \beta_B/\beta_A=0.9 'Ì"äŠr'¨'í'è
%%%%
%%\omega_B=0.8, \beta_A=4, \beta_B/\beta_A=1.1 'Ì"äŠr
%\begin{minipage}{0.5\hsize}
%\begin{center}
%\includegraphics[width=70mm]{HOg01_08_4_11.eps}
%\end{center}
%%\caption{First}
%%\label{fig.1}
%\end{minipage}
%\begin{minipage}{0.5\hsize}
%\begin{center}
%\includegraphics[width=70mm]{HOg001_08_4_11.eps}
%\end{center}
%%\caption{Caption two.}
%%\label{fig.2}
%\end{minipage}
%%\omega_B=0.8, \beta_A=4, \beta_B/\beta_A=1.1 'Ì"äŠrI'í'è
%%%
%%\omega_B=1.25, \beta_A=4, \beta_B/\beta_A=0.9 'Ì"äŠr
%\begin{minipage}{0.5\hsize}
%\begin{center}
%\includegraphics[width=70mm]{HOg01_125_4_09.eps}
%\end{center}
%%\caption{First}
%\end{minipage}
%\begin{minipage}{0.5\hsize}
%\begin{center}
%\includegraphics[width=70mm]{HOg001_125_4_09.eps}
%\end{center}
%%\caption{Caption two.}
%\end{minipage}
%%\omega_B=1.25, \beta_A=4, \beta_B/\beta_A=0.9 'Ì"äŠrI'í'è
%%%
%%\omega_B=1.25, \beta_A=4, \beta_B/\beta_A=1.1 'Ì"äŠr
%\begin{minipage}{0.5\hsize}
%\begin{center}
%\includegraphics[width=70mm]{HOg01_125_4_11.eps}
%\end{center}
%%\caption{First}
%\end{minipage}
%\begin{minipage}{0.5\hsize}
%\begin{center}
%\includegraphics[width=70mm]{HOg001_125_4_11.eps}
%\end{center}
%%\caption{Caption two.}
%\end{minipage}
%%\omega_B=1.25, \beta_A=4, \beta_B/\beta_A=1.1 'Ì"äŠrI'í'è
	\caption	
	{
		The quantities $\langle {\rm e}^{-\Delta \beta Q} \rangle_{\tau}-1$ 
		(solid red line) 
		and $\langle Q\rangle_{\tau}$ 
		(broken blue line) in the coupled harmonic oscillators. 
		We fixed $\beta_A \hbar \omega_A=4$: 
		for 
		$\omega_B / \omega_A	=0.8$, 
		$\beta_B / \beta_A		=0.9$ 
		with 
		(a)
		$\gamma \nu / \hbar \omega_A		=0.1$ 
		and 
		(b) 
		$\gamma \nu / \hbar \omega_A		=0.01$; 
		for  
		$\omega_B / \omega_A	=0.8$, 
		$\beta_B / \beta_A		=1.1$ 
		with 
		(c)
		$\gamma \nu / \hbar \omega_A		=0.1$ 
		and 
		(d) 
		$\gamma \nu / \hbar \omega_A		=0.01$; 
		for  
		$\omega_B / \omega_A	=1.25$,  
		$\beta_B / \beta_A		=0.9$ 
		with 
		(e) 
		$\gamma \nu / \hbar \omega_A		=0.1$ 
		and 
		(f) 
		$\gamma \nu / \hbar \omega_A		=0.01$; 
		for  
		$\omega_B / \omega_A	=1.25$,  
		$\beta_B / \beta_A		=1.1$ 
		with 
		(g) 
		$\gamma \nu / \hbar \omega_A		=0.1$ 
		and 
		(h) 
		$\gamma \nu / \hbar \omega_A		=0.01$. 
		}
\label{Fig:HO}
\end{figure}

To the second order of the coupling strength $\gamma$, 
the deviation from the integral exchange fluctuation theorem, 
$\langle {\rm e}^{-\Delta \beta Q}\rangle_{\tau}-1$,
and the heat transfer $\langle Q \rangle_{\tau}$ are analytically calculated 
as follows:
%		HO'Å'ÌIXFTƒYƒŒ'Æ"M'ÌŠú'Ò'l@ƒKƒ"ƒ}'Ì'QŽŸ'Ü'Å
\begin{align}
\langle e^{-\Delta \beta Q} \rangle_{\tau} -1
=&
	4\left\{
		\frac{\gamma \nu}{\hbar}
		\frac{\sin \left[ \frac{\tau}{2}(\omega_A -\omega_B ) \right]}%
{(\omega_A -\omega_B)/2}
	\right\}^2 
	\nonumber \\
	&\times
	\frac{\sinh \left(\Delta \beta \hbar \omega_A / 2 \right)
	\sinh \left[\beta_B \hbar ( \omega_A - \omega_B ) /2 \right]}%
{	\left[2 \sinh \left(\frac{\beta_A \hbar \omega_A}{2}\right) \right]
		\left[2 \sinh \left(\frac{\beta_B \hbar \omega_B}{2}\right) \right]
	}
	+\mathrm{O} (\gamma^3), \\
%\end{align}
%		Y
%		"MŒðŠ·
%\begin{align}
\langle Q \rangle_{\tau}
=&
	-2\left\{
		\frac{\gamma \nu}{\hbar}
		\frac{\sin \left[ \frac{\tau}{2}(\omega_A  -\omega_B ) \right]}%
{(\omega_A  -\omega_B )/2}
	\right\}^2 
	\nonumber \\
	&\times
	\frac{\hbar \omega_A
	\sinh \left[(\beta_A \hbar \omega_A- \beta_B \hbar \omega_B)/2\right]}%
{	\left[2 \sinh \left(\frac{\beta_A \hbar \omega_A}{2}\right) \right]
		\left[2 \sinh \left(\frac{\beta_B \hbar \omega_B}{2}\right) \right]
	}
	+\mathrm{O} (\gamma^3). 
\end{align}
%		Y
The $\tau$ dependence of the deviation from 
the integral exchange fluctuation theorem and the net heat transfer during $\tau$ 
are shown in Fig.~\ref{Fig:HO}. 
%
%'ቷ \beta 'å@'̏ꍇ'ðŽ¦'µ'Ä'à'¢'¢'ÆŽv'¤'ª'Ç'¤'¾'낤'©B
%%%%%%%%%%%%%%%%%%
%%%%%%%%%%%%%%
%%%%%%%%%%%
%%%%%%%%
%%%%%
%%
%			
%		ƒpƒ‰ƒƒ^'̏‡"Ô		\omega_B,  \beta_A,  \gamma \beta=\beta_B//beta_A
%%%
%			HO'̃Oƒ‰ƒt
%	\nu \gamma=1,	\omega_B=0.8,		\beta_A=4,	\gamma\beta=0.9
%%
%%%%%%%%%%%		TwoSpin'ÌXFTƒYƒŒ'Æ"M—¬
%						HO'̐}'Í'¨I'¢B
%%%%
%%%%%%
%%%%%%%%%%
%%%%%%%%%%%%%%%
To the second order of the coupling strength $\gamma$, 
the ratio of the above two quantities is given by 
%XFT'Æ"M—¬'Ì"ä
\begin{align}
\frac{\langle {\rm e}^{-\Delta \beta Q} \rangle_{\tau}-1}
	{\langle Q \rangle_{\tau}/(\frac{\hbar \omega_A}{2})}
		&=
		\frac
		{\sinh \left(\Delta \beta \hbar \omega_A / 2 \right)
		\sinh \left[ \beta_B \hbar(\omega_A-\omega_B)/2 \right]
		}		
		{
		\sinh \left[ \hbar(\beta_B \omega_B - \beta_A \omega_A)/2 \right]
		}
		+\mathrm{O}(\gamma^3).
\label{RatioHO}
\end{align}
%		Y
This result again shows that the exchange fluctuation theorem does not generally hold 
in the presence of a finite heat transfer. 

However, we can see that 
the integral fluctuation theorem holds in the presence of a finite heat transfer 
at the point given by $\omega_{A}=\omega_{B}$. 
At this point and at this point only, 
the commutable-coupling condition is satisfied as 
%	HO'ÅCCC
\begin{align}
[H_0, H_{\rm c}]
	=
		\hbar \nu (\omega_A - \omega_B)
		(a^{\dagger}b - b^{\dagger}a)
	=0.
%\label{TwoHOCCC}
\end{align}
%		Y

For the more general coupling Hamiltonian, 
$H_{\rm c}=\nu \left[ (a^{\dagger})^{l_A} b^{l_B}  +  (b^{\dagger})^{l_B} a^{l_A} \right]$, 
the deviation from the integral exchange fluctuation theorem and the net heat transfer 
are calculated to the second order of $\gamma$ as 
%		HO'Å'ÌIXFTƒYƒŒ'Æ"M'ÌŠú'Ò'l@ƒKƒ"ƒ}'Ì'QŽŸ'܂Ł@ˆê"Ê'Ì l_A
\begin{align}
\langle e^{-\Delta \beta Q} \rangle_{\tau} -1
=&
	\left\{
		\frac{\gamma \nu}{\hbar}
		\frac	{\sin \left[ \frac{\tau}{2}(\omega_A l_A -\omega_B l_B) \right]}
		{(\omega_A l_A -\omega_B l_B)/2}
	\right\}^2 
\nonumber\\
&\times
	\frac{4 l_A !\ l_B !}
	{	\left[2 \sinh \left(\beta_A \hbar \omega_A/2\right) \right]^{l_A}
		\left[2 \sinh \left(\beta_B \hbar \omega_B/2\right) \right]^{l_B}
	}
	\nonumber \\
	&\times
	\sinh \left(\Delta \beta \hbar \omega_A \l_A / 2 \right)
	\sinh \left[\beta_B \hbar ( \omega_A l_A - \omega_B l_B) /2 \right]
		+\mathrm{O}(\gamma^3), 
\end{align}
%		Y
%		"MŒðŠ·
\begin{align}
\langle Q \rangle_{\tau}
=&
	\left\{
		\frac{\gamma \nu}{\hbar}
		\frac	{\sin \left[ \frac{\tau}{2}(\omega_A l_A -\omega_B l_B) \right]}
		{(\omega_A l_A -\omega_B l_B)/2}
	\right\}^2 
\nonumber\\
&\times
	\frac{-2 l_A !\ l_B !}
	{	\left[2 \sinh \left(\beta_A \hbar \omega_A/2\right) \right]^{l_A}
		\left[2 \sinh \left(\beta_B \hbar \omega_B/2\right) \right]^{l_B}
	}
	\nonumber \\
	&\times
	\hbar \omega_A l_A
	\sinh \left[(l_A \beta_A \hbar \omega_A-l_B \beta_B \hbar \omega_B)/2\right]
		+\mathrm{O}(\gamma^3). 
\end{align}
%		Y
The ratio of the above two quantities is given by 
%XFT'Æ"M—¬'Ì"ä	l_Aˆê"Ê
\begin{align}
\frac{\langle {\rm e}^{-\Delta \beta Q} \rangle_{\tau}-1}
	{\langle Q \rangle_{\tau}/(\frac{\hbar \omega_A l_A}{2})}
		&=
		\frac
		{\sinh \left(\Delta \beta \hbar \omega_A l_A / 2 \right)
		\sinh \left[ \beta_B \hbar(\omega_A l_A-\omega_B l_B)/2 \right]
		}		
		{
		\sinh \left[ \hbar(\beta_B \omega_B l_B - \beta_A l_A \omega_A)/2 \right]
		}
		+\mathrm{O}(\gamma^3). 
\label{RatioHOGeneral}
\end{align}
%		Y
This shows that the integral 
exchange fluctuation theorem holds in the presence of a finite heat 
transfer if and only if the commutable-coupling condition is satisfied: 
%	CCC in HO@l_Aˆê"Ê
\begin{align}
[H_0, H_{\rm c}]
	=
		\hbar \nu (\omega_A l_A - \omega_B l_B)
		\left[(a^{\dagger})^{l_A}b^{l_B} - (b^{\dagger})^{l_B}a^{l_A}\right]
	=0,
%\label{TwoHOCCC}
\end{align}
%		Y
or $\omega_A l_A = \omega_B l_B$.

%Œ‹˜_
%\newpage
\section{Conclusions} \label{Chap:Conclusion}
To summarize, we showed that the exchange fluctuation theorem 
in its original form 
%ƒIƒŠƒWƒiƒ‹'ÌŽ®'ÉŽ'Á'Ä'­'é'Æ'«'ÉŽžŠÔ"½"]••Õ«'ÌŒø'«•û'É'æ'Á'Ă̓IƒŠƒWƒiƒ‹'ÆŠ®'S'Ɉê'v'·'é'Ì'¾'ªB
does not generally hold 
in the presence of a finite heat transfer. In the limit $\gamma \to 0$, the $k$th moments of $p_{\tau}(Q)$ vanish for $k \geq 1$. 
The deviation from the exchange fluctuation theorem 
also vanishes in the same order of $\gamma$. 
We derived general formulas for the above and analytically confirmed them 
for specific models. 
This means that there is no heat transfer 
when the coupling strength $\gamma$ is small enough to neglect the deviation from the exchange fluctuation theorem. 
In this case, the exchange fluctuation theorem reduces to a trivial relation and has no information about the heat transfer. 

However, 
we found a condition for the exchange fluctuation theorem to hold exactly 
for arbitrary $\gamma$ and we referred to it as the commutable-coupling condition. 
Under this condition, the exchange fluctuation 
theorem becomes an exact relation independently of the coupling strength $\gamma$ 
under the existence of a finite heat transfer.
We confirmed this in specific models. 
In short, the exchange fluctuation theorem holds not because the coupling is weak 
as was originally proposed, but because the total energy of the system is conserved 
under the commutable-coupling condition.

The deviation from the exchange fluctuation theorem consists of  the commutation relation between the Hamiltonian of 
the total system and the coupling Hamiltonian. 
Therefore, the non-commutativity of the observable in quantum 
mechanics plays an important role in the deviation. 

%\newpage
\section*{Acknowledgment}
We are grateful to Professor Hisao Hayakawa 
for his valuable comments and suggestions. 
This work is supported by Grant-in-Aid for Scientific Research No.~17340115 
from the Ministry of Education, Culture, Sports, Science and Technology as well as 
by Core Research for Evolutional Science and Technology (CREST) of Japan 
Science and Technology Agency.

%\section*{Acknowledgements}
%We would like to thank ...........

%\appendix
%\section{First Appendix} %Empty argument \section{} yields `Appendix'. 
%
%\section{Second Appendix}

%ŠÖ˜A˜_•¶@ŽQl•¶Œ£


\begin{thebibliography}{99}
%%%%%%%%%%%%%%%%%%%%%%%%%%%%%%%%%%%%%%%%%%%%%%%%%%%%%%%%%%%%%
% Some macros are available for the bibliography:
%  o for general use
%    \JL : general journals                 \andvol : Vol (Year) Page
%  o for individual journal 
%    \AJ   : Astrophys. J.           \NC         : Nuovo Cim.
%    \ANN  : Ann. of Phys.           \NPA, \NPB  : Nucl. Phys. [A,B]
%    \CMP  : Commun. Math. Phys.     \PLA, \PLB  : Phys. Lett. [A,B]
%    \IJMP : Int. J. Mod. Phys.      \PRA - \PRE : Phys.~Rev.\ [A-E]     
%    \JHEP : J. High Energy Phys.    \PRL        : Phys.~Rev.\ Lett.
%    \JMP  : J. Math. Phys.          \PRP        : Phys. Rep.
%    \JP   : J. of Phys.             \PTP        : Prog. Theor. Phys.     
%    \JPSJ : J. Phys. Soc. Jpn.      \PTPS       : Prog. Theor. Phys. Suppl.
% Usage:
%  \PRD{45,1990,345}          ==> Phys.~Rev.\ D \textbf{45} (1990), 345
%  \JL{Nature,418,2002,123}   ==> Nature \textbf{418} (2002), 123
%  \andvol{123,1995,1020}    ==> \textbf{123} (1995), 1020
%%%%%%%%%%%%%%%%%%%%%%%%%%%%%%%%%%%%%%%%%%%%%%%%%%%%%%%%%%%%%
%FT'̍ŏ‰'Ì'±'ë		"™‰·"M—Œn
	\bibitem{FToriginal1993:Evans}
	D. J. Evans, E. G. D. Cohen, and G. P. Morriss, 
	Phys.~Rev.\ Lett. \textbf{71} (1993), 2401.

	\bibitem{FToriginal1994:Evans}
	D. J. Evans, D. J. Searles, 
	Phys.~Rev.\ E. \textbf{50} (1994), 1645.

	\bibitem{FT1995:GallavCohen}
	G. Gallavotti and E. G. D. Cohen,
	Phys.~Rev.\ E. \textbf{74} (1995), 2694.

%FT	Šm—¦‰ß'ö			ŒÃ"T
	\bibitem{FTStocha1998:Kurchan}
	J. Kurchan,
	J. Phys. A: Math. Gen. \textbf{31} (1998), 3719.

	\bibitem{FTStocha1999:Lebow}
	J. L. Lebowitz and H. Spohn,
	J. Stat. Phys. \textbf{95} (1999), 333.


%FT	ŽdŽ–'ÉŠÖ'·'é		ŒÃ"TŒn
	\bibitem{JE1997:Jarzyn}
	C. Jarzynski,
	Phys.~Rev.\ Lett. \textbf{78} (1997), 2690.

	\bibitem{FTw1998:Crooks}
	G. E.  Crooks,
	J. Stat. Phys. \textbf{90} (1998), 1481.

	\bibitem{FTw1999:Crooks}
	G. E.  Crooks,
	Phys.~Rev.\ E. \textbf{60} (1999), 2721.

%	\bibitem{FTw2000:Crooks}
%	G. E.  Crooks,
%	Phys.~Rev.\ E. \textbf{60}, 2361 (2000).

%ƒŒƒrƒ…[iŒÃ"T'Ì—h'炬'̒藝j
	\bibitem{RevFT2002:Evans}
	D. J. Evans and D. J. Searses,
	Adv. Phys.  \textbf{51} (2002), 1529. 

	\bibitem{RevFT2008:Searles}
	E. M. Sevick, R. Prabhakar, S. R. Williams, and D. J. Searles, 
	Annu. Rev. Phys. Chem. \textbf{59} (2008), 603. 


%TFT'ɑ΂·'éŽÀŒ±"IŒŸØ()

%SSFT'ɑ΂·'éŽÀŒ±"IŒŸØ
	\bibitem{ExpSSFT2004:Carberry}
	D. M. Carberry, J. C. Reid, G. M. Wang, E. M. Sevick, D. J. Searles, and D. J. Evans,
	Phys.~Rev.\ Lett. \textbf{92} (2004), 140601.

%JE@'ɑ΂·'éŽÀŒ±"IŒŸØ
	\bibitem{ExpJE2002:Liphardt}
	J. Liphardt, S. Dumont, S. B. Smith, I. Tinoco (Jr), and C. Bustamante,
	Science \textbf{296} (2002), 1832. 

%Crooks@'ɑ΂·'éŽÀŒ±"IŒŸØ
	\bibitem{ExpFTw2005:Collin}
	D. Collin, F. Ritort, C. Jarzynski, S. B. Smith, I. Tinoco (Jr), and C. Bustamante,
	Nature \textbf{437} (2005), 231.


%	—ÊŽq‰‰ŽZŽq'ÅŽdŽ–'ð'è‹`
	\bibitem{OpeWork2000:Yukawa}
	S. Yukawa, 
	J. Phys. Soc. Jpn. \textbf{69} (2000), 2367.
	
	\bibitem{OpeWork2003:Monnai}
	T. Monnai and S. Tasaki,
	cond-mat/0308337.
	
	\bibitem{OpeWorkHeat:Allahverdyan}
	A. E. Allahverdyan and Th. M. Nieuwenhuizen,
	Phys.~Rev.\ E \textbf{71} (2005), 066102.

	\bibitem{OpeWorkHeat:GelinKosov}
	M. F. Gelin and D. S. Kosov,
	Phys.~Rev.\ E \textbf{78} (2008), 01116.
	
	
%	Two-time measurement schem
	\bibitem{TwoTime:Kurchan}
	J. Kurchan, 
	cond-mat/0007360.
	
	\bibitem{TwoTime:HalTasaki}
	H. Tasaki, cond-mat/0009244.

	\bibitem{TwoTime:Mukamel}
	S. Mukamel, 
	Phys.~Rev.\ Lett. \textbf{90} (2003), 170604.
	
	\bibitem{TwoTime:Monnai}
	T. Monnai,
	Phys.~Rev.\ E. \textbf{72} (2005), 027102.

	\bibitem{TwoTime:TalknerWorkNot}
	P. Talkner, E. Lutz, and P. Hanggi,  
	Phys.~Rev.\ E. (R)\textbf{75} (2007), 050102.


%XFT
	\bibitem{JarzynWo}
	C. Jarzynski and D. K. W\'{o}jcik, 
	Phys.~Rev.\ Lett. \textbf{92} (2004), 230602.
	
	
\if0		
%	‹ï'Ì"I'ȃ'ƒfƒ‹'ÅSSFT'ðØ–¾
	\bibitem{SSFT2007:Saito}
	K. Saito and A. Dhar, 
	Phys.~Rev.\ Lett. \textbf{99} (2007), 180601.
	
	\bibitem{SSFT2008:SaitoU}
	K. Saito and Y. Utsumi, 
	Phys.~Rev.\ B. \textbf{78} (2008), 115429. 
	
%	SSFT'Ì—ÊŽqŒn'ł̏ؖ¾
	\bibitem{SSFTproof2008:Andrieux}
	D. Andrieux, P. Gaspard, T. Monnai, S. Tasaki
	cond-mat/08113687. 
\fi	

%ŽžŠÔ"½"]‰‰ŽZŽq
	\bibitem{TRQuant1}
	J. J. Sakurai, \textit{Modern Quantum Mechanics}, 
	(Benjamin, Menlo Park, California, 1985). 

%CCC‰º'ŃWƒƒƒ‹ƒ`ƒ"ƒXƒL['ð‹c˜_
	\bibitem{CCCpaper}
	J. Teifel and G. Mahler,
	Phys.~Rev.\ E \textbf{76} (2007), 051126. 

\if0	
%	‹t‰ß'ö'É'¨'¯'éƒWƒ‡ƒCƒ"ƒgŠm—¦
	\bibitem{RJointP2009:Esposito}
	M. Esposito, U. Harbola, and S. Mukamel, 
	cond-mat/08113717. 
\fi

\if0
	%Ž¥ê"ü'è'Å'à‹t‰ß'ö'ÌŽžŠÔ"­"W‰‰ŽZŽq'ð'è‹`B
	\bibitem{RTEvo2008:Andrieux}
	D. Andrieux and P. Gaspard
	Phys.~Rev.\ Lett. \textbf{100} (2008), 230404.
\fi

\end{thebibliography}
\end{document}